\documentstyle[12pt]{article}
\begin{document}

\title{Fractalization of Torus Revisited as a Strange Nonchaotic Attractor}

\author{
        Takashi Nishikawa and Kunihiko Kaneko\\
        {\small \sl Department of Pure and Applied Sciences}\\
        {\small \sl University of Tokyo, Komaba, Meguro-ku, Tokyo 153, JAPAN}\\
\\}

\date{}
\maketitle
\begin{abstract}
Fractalization of torus and its transition to chaos in a quasi-periodically
forced logistic map is re-investigated in relation with a strange
nonchaotic attractor, with the aid of functional equation for the invariant
curve.  Existence of fractal torus in an interval in parameter space is
confirmed by the length and the number of extrema of the torus attractor,
as well as the Fourier mode analysis.  Mechanisms of the onset of fractal
torus and the transition to chaos are studied in connection with the
intermittency.
\end{abstract}

\section{Introduction}
\mbox{}

Transition from a torus to chaos has been intensively
investigated\cite{KK-Book}. There two types of instabilities exist, which
cause the collapse of tori: One is the instability along the phase
direction and the other in the amplitude direction.  The former instability
has already been studied in detail using the circle map while the
oscillation and fractalization of tori have been reported due to the
instability along the amplitude direction.  One of the authors (KK)
reported that the oscillation of tori gets stronger with the increase of
external forcing, until it reaches fractal, and then chaos appears
\cite{KK}.  Although the fractal nature of the torus was confirmed at the
onset of chaos, the strange oscillation of torus before the onset remained
unclear.

On the other hand the existence of strange nonchaotic attractors(SNA) was
shown by Grebogi et al\cite{GOY}.  Here the word ``strange'' refers to the
geometry of the attractors, and the word ``chaotic'' refers to the orbital
instability of the dynamics.  In a model with quasiperiodic forcing
similar to that for the fractalization of torus, they showed analytically
that the attractor has a non-positive Lyapunov exponent but a complicated
geometry.  Indeed the attractor is not smooth on the set of dense points. Although the
original model by Grebogi et al. \cite{GOY} excludes the possibility of
chaos, the SNA has generally been observed in a system where chaos
appears with the increase of the forcing as was discussed in the
fractalization of torus \cite{PF,FKP,HH,Ani,Ding,Lai}.

In the present paper, we study how the transition from smooth to fractal
torus occurs. We have confirmed that there is a parameter region with a
non-zero measure where the SNA exists between the smooth torus and
chaos. This leads to the following scenario of the destruction of tori.
\begin{center}
Torus$\rightarrow$Fractal Torus(SNA)$\rightarrow$Chaos
\end{center}

In the following sections, we will first characterize the nature of fractal
torus as a SNA.  Besides the direct simulation of the quasi-periodically
forced logistic map, we adopt the functional map to obtain the invariant
torus.  The length of the invariant curve, as well as the number of
singular points, shows how the fractalization occurs with the parameter
change.  In \S 3, perturbation expansion of the functional
equation is given, where the SNA is characterized by the loss of the
convergence in the Fourier mode expansion of the derivative of the
invariant curve.  Comparison between smooth and fractal tori is given in \S
4, where some dynamical signatures of the fractal torus is discussed, such
as the parameter sensitivity of the Lyapunov exponent, and the phase
sensitivity.  The onset of chaos from the SNA is investigated in \S 5,
where the transition is associated with the intermittency from the fixed
point in the functional map.

\section{Fractalization of Torus as SNA}
\mbox{}

Transition from torus to chaos has been studied with a variety of
2-dimensional maps\cite{KK-Book}.  With the change of the bifurcation parameter, the
amplitude of oscillation gets stronger, accompanied with phase-lockings.  To
focus on the amplitude instability by excluding the phase locking, we
choose the following two dimensional map \cite{KK}, with a constant
rotation of the phase;

\begin{equation}
\{
\begin{array}{ll}
x_{n+1}=f(x_n)+\varepsilon g(\theta_n)\\
\theta_{n+1}=\theta_n+\omega \bmod 1
\label{MAP}
\end{array}
\end{equation}

Here $x_n$ represents the amplitude, while $\theta_n$ corresponds to the
phase of oscillation. To exclude the phase lockings no coupling from $x$ to
$\theta$ is included, where $\omega$ is an irrational number representing
the rotation number.

In this paper we set $f(x)=ax(1-x)$,$g(\theta )=\sin 2\pi \theta$, and
$\omega=(\sqrt{5}-1)/2$, as the simplest nonlinear map with an external
driving force.  For most simulations, we fix $a=3.0$, and vary
$\varepsilon$ to see the change of the attractor. At $\varepsilon=0$, the
map, of course, is just a one-dimensional map of $x_n$ with a fix point
$x^*=1-1/a$. Thus the attractor of the two-dimensional map is just a 
with a straight line torus.  As $\varepsilon$ is increased, oscillations of the
torus start to appear, which become fractal with the dimension above one (
as is confirmed later by direct measurement of the length of the torus).
As $\varepsilon$ is increased further, another transition from fractal
torus to chaos appears, by which the stability along the $x$-direction is
lost.
 
\begin{figure}
\begin{center}
\mbox{}
\caption{Three types of attractor: The value of $\varepsilon$ in each
figure is (a)$\varepsilon =0.1$, (b)$\varepsilon =0.156$, (c)$\varepsilon =0.18$,
respectively. We have plotted 30,000 points after discarding initial
transients, where $a=3.0, \omega =(\protect\sqrt{5}-1)/2$. }
\label{ATTR}
\end{center}
\end{figure}

\begin{figure}
\caption{Lyapunov exponents are plotted versus the parameter $\varepsilon$. 
Computed from the average over $10^6$ iterations of the map (1), after
discarding initial 5,000 points of transient.}
\label{LYAP}
\end{figure}

Three examples of the patterns of attractors are given in Fig.\ref{ATTR}
corresponding to the three types of attractors, while the change of the
Lyapunov exponent is plotted in Fig.\ref{LYAP}.  Transition from torus to
chaos occurs at $\varepsilon \sim 0.1573$. As will be confirmed later, the
torus loses its smoothness around $\varepsilon \sim 0.1553$. Hence the SNA
exists in an interval of the parameter, where the Lyapunov exponent has a
sharp sensitivity to $\varepsilon$, as in the chaos region.

To confirm that the attractor has a non-integer dimension, we study
the equation for the invariant torus, following the argument of
\cite{KK}. In our model (\ref{MAP}) the attractor is expressed as a
single-valued function of $\theta$, as $x=X(\theta )$ ( $0 \le \theta
\le 1$ ). If this function represents an invariant curve of the map,
it must satisfy the following functional equation.

\begin{eqnarray}
X(\theta + \omega ; \bmod 1) &=& f(X(\theta ))+\varepsilon g(\theta )\nonumber\\
&=& aX(\theta )(1-X(\theta))+\varepsilon \sin 2\pi \theta 
\label{FUNC_EQ}
\end{eqnarray}

This equation is postulated by the constraint that the point
$(x_{n+1},\theta_{n+1})$ should be also on the curve, and is obtained by
substituting $x_n$ by $X(\theta )$ and $x_{n+1}$ by $X(\theta + \omega )$
in the map (\ref{MAP}). From this functional equation we can introduce the
following ``functional map", which maps a function to another function:

\begin{eqnarray}
X_{n+1}(\theta + \omega ; \bmod 1) &=& f(X_n(\theta ))+\varepsilon g(\theta )
\nonumber\\
&=& aX_n(\theta )(1-X_n(\theta))+\varepsilon \sin 2\pi \theta .
\label{FUNC_MAP}
\end{eqnarray}

The attractor of the original map (\ref{MAP}) is obtained as a fixed point
in the {\it functional space}, for the iteration of the functional map
(\ref{FUNC_MAP}). This discussion stands only if the Lyapunov exponent is
negative and the attractor has a single value $x$ for each $\theta$.

Since the functional map (\ref{FUNC_MAP}) is an infinite dimensional
dynamical system, one cannot compute it directly.  We have numerically
computed it by approximating $\omega=(\sqrt{5}-1)/2$ by
$\omega_k=F_{k-1}/F_k$, where $\{F_k\}_{k=0,1,...}$ are Fibonacci
series. ($\omega_k \rightarrow \omega$ as $k \rightarrow \infty$ ). This
approximation transforms the functional map to an $F_k$-dimensional map
which maps $F_k$ lattice points on the $\theta$ coordinate onto themselves. 
We have computed the attractor of this $F_k$-dimensional map, to obtain the
approximate solution of the functional equation(\ref{FUNC_EQ}) as a
piecewise-linear function.

For the parameter regime corresponding to the torus attractor, the
convergence is rather fast.  The convergence time of the functional map to
a fixed point gets longer as $\varepsilon$ approaches the onset of chaos,
while, for $\varepsilon >0.1573$, corresponding to the chaos region, the
functional map does not converge to a fixed point.

The functional equation enables us to compute the length $L_j$ of $X(\theta
)$ as follows:

\begin{equation}
L_j = \sum_{j=0}^{F_k-1}\sqrt{\left\{X\left(\frac{i}{F_k}\right)-X\left(\frac{i+j}{F_k}\right)\right\}^2+\left(\frac{j}{F_k}\right)^2}
\label{L_j}
\end{equation}

If the dimension of the attractor is $1+\alpha$, $L_j$ must scale as
$L_j\propto j^{-\alpha }$, where $0<\alpha<1$. The slope of this plot
$\alpha$ gives a kind of fractal dimension. Fig.\ref{LENGTH} is an example
of the log-log plot of length $L_j$ versus the mesh width $j$.
Here we have adopted $F_k=317811$ for the approximation of $\omega$. Hence
$j=1$ corresponds to the mesh size of $1/317811 \sim 3\times 10^{-6}$.
Fig.\ref{FRACDIM3.0} is the plot of slope $\alpha$ versus external force
$\varepsilon$.

\begin{figure}
\caption{Examples of log-log plot of length versus mesh
($a=3.0,F_k=317811$), obtained from the attractor of the functional map.}
\label{LENGTH}
\end{figure}

\begin{figure}
\caption{Change of the slope $\alpha$ versus external forcing $\varepsilon$
($a=3.0,F_k=317811$).  The slope is estimated from the length at the 5
smallest mesh scales.  The error bar is not explicitly given, but it is
about $.05 \sim .1$.}
\label{FRACDIM3.0}
\end{figure}

The slope starts to be nonzero at the transition point at about
$\varepsilon \sim 0.1548$. The slope jumps to $\alpha \sim 0.62$, and stays
around the value with the increase of $\varepsilon$, up to the vicinity of
the onset of chaos.

\begin{figure}
\caption{Length of attractors with the mesh scale $j$. $F_k=9227465$, and
$a=3.0$.}
\label{FINE_MESH}
\end{figure}

In the region between $\varepsilon \approx .1548$ and $.1553$, the exponent
is roughly 0.3 up to fine mesh scales (e.g., $1/317811$), but there appears
a saturation at a finer mesh.  In Fig.\ref{FINE_MESH} we have plotted the
length versus mesh, taking $F_k=9227465$. There is a crossover to a smooth
behavior at the scale about $10/9227465$ for $\varepsilon=0.1548$.  Here we
call this parameter regime as ``pre-fractal-torus'', since this region is
distinguished from the smooth and fractal torus regimes.  Indeed the
crossover to a smooth curve is seen only at a much finer scale than
for a smooth torus regime, while this crossover scale increases as
$\varepsilon$ approaches the onset parameter for fractal torus.  The slope
at the scaling regime (i.e., scales with a larger mesh beyond the crossover) has a
jump at the onset of fractal torus: The exponent $\alpha$ in the scaling
regime jumps from .3 to .62, when the parameter $\varepsilon$ crosses
$0.1553$, the border between pre-fractal and fractal torus regimes.

For $\varepsilon > 0.1573$, $X_n(\theta )$ does not converge to a fixed
curve.  Indeed the Lyapunov exponent of the two-dimensional map is positive
in this regime, and the attractor here is chaotic. Summing up the results
of the fractal exponent of Fig.\ref{FRACDIM3.0}, and the Lyapunov exponent
of Fig.\ref{LYAP}, one can conclude that the fractal torus exists at least
in the region $0.1553 < \varepsilon < 0.1573$ \footnotemark.

We have also measured the distribution of local dimensions, by varying
$\varepsilon$, which is shown in Fig.\ref{LOCAL}.  The distribution is
single-humped, with slight asymmetry.  At the ``pre-fractal-torus'', the
distribution has a peak around zero.

\begin{figure}
\caption{Distribution of the local dimension obtained through the measurement of the local
length. We have measured the length of $X(\theta)$ at each lattice point
using 6000 lattice points around it. By computing the local dimension over
all lattice points, and sampling the number with the bin size 0.01, the
distribution is obtained.  ($a=3.0$)}
\label{LOCAL}
\end{figure}

\footnotetext{By numerical means, one cannot completely exclude the
possibility that the the non-smoothness of the torus disappears as the mesh
scale goes down to a much finer scale, and the fractal torus exists only at
the onset of chaos.  Indeed this suspect has lead to the report of the
fractal torus only at a single parameter point, in \cite{KK}. We assume
that the scaling with the exponent $\alpha$ lasts to an infinitesimal
scale, since the increase of the crossover size near $0.1553$ is so strong
that it is distinguished from the pre-fractal-torus region.}

We have also calculated the number of extremal points of the function
$X(\theta )$.  Here the number $N_j$ of extremal points for a given mesh is
defined as follows. First count the number of the mesh point $i$ such that

\begin{equation}
\left\{X\left(\frac{i+j}{F_k}\right)-X\left(\frac{i}{F_k}\right)\right\} 
\left\{X\left(\frac{i+2j}{F_k}\right)-X\left(\frac{i+j}{F_k}\right)\right\} <0
\end{equation}
is satisfied: Then $N_j$ is obtained by dividing this number by
$j$. Fig.\ref{EXTREMA} gives some examples of the log-log plot of $N_j$
versus $j$.

\begin{figure}
\caption{The number of extremal points versus mesh size.  See text for the
method of computation.  $(a=3.0,F_k=317811)$.}
\label{EXTREMA}
\end{figure}
 
The number $N_j$ increases as the mesh $j$ is reduced, which means that the
finer mesh one uses, the more extrema appear. This is consistent with the
fractal nature of torus.  However, we note that $N_j$ scales not as $N_j
\sim j^{-1}$ but as $N_j \sim j^{-\beta}$, where $0<\beta<1$. This means
that the fraction of singular points decreases with the mesh.  In other words,
the extremal points lie on a Cantor set on the $\theta$ axis.  

Dependence of $\beta$ on the parameter $\varepsilon$ is plotted
in Fig.\ref{FRACDIM3.0}.  The exponent stays around 0.75 while the fractal
dimension stays around 0.6.  Near the onset of chaos, $\varepsilon
=0.1573$, the exponent $\beta$ increases, till it takes 1.0 at the
onset.  In other words, the ruggedness of the torus fills the space down to
any small scale when the torus loses stability and is replaced by chaos.

The linear increase of the extremum point agrees with the picture by a
random curve.  Consider, for example, a curve $C(i)$ (with $i$ lattice
point) generated by $C(i+1)=\delta \times C(i)+\eta$ with $\eta$ a random
number distributed over [-1,1] and $\delta \leq 1$.  The number of extremum
points increases linearly with the number of mesh points.  The increase is
given by $2N/3 $ for $\delta =0$, while the increase is slower with a
smaller proportion coefficient for $\delta >0$, till it is given
by $N/2$ for $\delta=1$.  Since the increase is given by $0.58N$ at the
onset of chaos (and also in the chaos region) in our simulation, the
critical torus can be approximated by a correlated random curve.  This
behavior and the coefficient characterize how the fractal torus collapses
and is replaced by chaos in the two dimensional map eq.(\ref{MAP}).

To close the present section, it is interesting to note the universality of
our results.  In the discovery of a fractal torus by \cite{KK}, the
parameter $a=-1$ was adopted, where successive transition from torus to
fractal torus, and then to chaos was found with the increase of
$\varepsilon$.  We have re-examined the simulation using a larger mesh
sizes.  Here again, a fractal torus exists for a finite interval of
parameters, where the fractal dimension increases from 0.6 to about $0.88$
with $\varepsilon$.  The number of extremal points again increases with a
fractional power, while it increases linearly at the onset of chaos.

Another interesting set of examples is given for $a=2.8$.  The Lyapunov
exponent is plotted in Fig.\ref{LYAP2.8}, while the fractal dimension and
the number of extremal points obtained from the functional equation (with
the mesh 317811) are given in Fig.\ref{FRACDIM2.8}.

\begin{figure}
\caption{(a),(b)Change of the Lyapunov exponent with the parameter
$\varepsilon$($a=2.8,\omega=(\protect\sqrt{5}-1)/2$).  (b) is the blowup of
(a) for $0.204<\varepsilon<0.208$.  The exponent is computed from the
average of $10^6$ steps after discarding initial 5000 steps.}
\label{LYAP2.8}
\end{figure}

\begin{figure}
\caption{Fractal dimension $\alpha$ (obtained by measuring the length) and the
exponent $\beta$ on the number of extremal points are plotted versus $\varepsilon$.  
Computed from the attractor of the functional map.($a=2.8,F_k=317811$).}
\label{FRACDIM2.8}
\end{figure}

From these figures, one can see that there appears windows of fractal torus
(FT) and smooth torus(ST), beyond the onset of chaos.  Indeed there are
transition sequences among ST, FT, chaos(C), as is seen in
Fig.\ref{LYAP2.8} and in Fig.\ref{FRACDIM2.8}.  At $\varepsilon \sim .204$ FT
appears, which is replaced by chaos at $\varepsilon \sim .206$. For $\varepsilon
\sim .208$, the ``inverse" bifurcation from chaos to FT, and then to ST (at
$\varepsilon \sim .211$) proceeds. Further bifurcations to FT and chaos, and
back to ST are seen for larger $\varepsilon$.  Again at the boundary between
chaos and fractal torus, the increase of the extremum points is almost
linear.

\section{Fractal Torus viewed from Functional Map}

\mbox{}

In this section we apply the Fourier mode analysis 
to the functional equation(\ref{FUNC_EQ}). First we consider the
Fourier expansion of $X(\theta )$
\begin{eqnarray}
X(\theta )=\sum_{k=-\infty}^{\infty}\hat{X}(k)e^{2 \pi ik \theta }\label{FS}\\
\hat{X}(k)=\int_0^1 X(\theta )e^{-2 \pi ik \theta } d{\theta } 
\end{eqnarray}
and substitute it into (\ref{FUNC_EQ}). Then we have
\begin{equation}
\hat{X}(k)e^{2 \pi ik \omega }=a\hat{X}(k) -a\sum_{k'=-\infty}^{\infty} \hat{X}(k')\hat{X}(k-k') - 
\frac{i \varepsilon}{2}(\delta_{k,1} - \delta_{k,-1})
\label{*}
\end{equation}
Next we expand each Fourier mode $X(k)$ with respect to the powers of
$\varepsilon$.
\begin{equation}
\hat{X}(k) = \sum_{n=0}^{\infty} \varepsilon^n \hat{X}_n(k)
\end{equation}
Substituting this into (\ref{*}) and comparing the terms of same power of 
$\varepsilon$ in both sides of equation, we get 
\begin{equation}
\hat{X}_n(k)e^{2 \pi ik \omega } = a\hat{X}_n(k) - a\sum_{k'=-\infty}^{\infty}\sum_{m=0}
^n \hat{X}_m(k')\hat{X}_{n-m}(k-k') - \frac{i}{2}\delta_{n,1}(\delta_{k,1} - \delta_{k,-1})
\label{PER_EQ}
\end{equation}
for each $n$. It is straightforward to show that $\hat{X}(k)$ has only
terms $\varepsilon^n$ such that $n \ge |k|$. This means that the term
$\varepsilon^n$ only has a Fourier mode with the wave number no more than
$n$. The first three terms of $\hat{X}_n(k)$, $\hat{X}(k)$ and $X(\theta)$ are
given as follows:
\begin{eqnarray}
\hat{X}_0(k) &=& \left( 1-\frac{1}{a}\right)\delta_{k,0}\\
\hat{X}_1(k) &=& \frac{\pm i \delta_{k,\pm 1}}{2(e^{\pm 2 \pi i 
\omega} + a - 2)}\\
\hat{X}_2(k) &=& \frac{-a\{ 2\hat{X}_1(-1)\hat{X}_1(1)\delta_{k,0} + (\hat{X}_1(-1))^2\delta_{k,-2} + 
(\hat{X}_1(1))^2\delta_{k,2}\} }{e^{2 \pi ik \omega}+a-2}\\
\nonumber\\
\hat{X}(0)&=&\hat{X}_0(0)+\varepsilon^2 \hat{X}_2(0)+\varepsilon^4 \hat{X}_4(0)+ \cdots \\
\hat{X}(\pm1 )&=&\varepsilon \hat{X}_1(\pm 1) +\varepsilon^3 \hat{X}_3(\pm 1) +\varepsilon^5 \hat{X}_5(\pm 1)+\cdots\\
\hat{X}(\pm 2)&=&\varepsilon^2 \hat{X}_2(\pm 2) +\varepsilon^4 \hat{X}_4(\pm 2)+\cdots\\
&\ldots&\nonumber\\
\nonumber\\
X(\theta)&=&\hat{X}_0(0)+\varepsilon \left\{ \hat{X}_1(1)e^{2 \pi i \theta}+\hat{X}_1(-1)e^{-2 \pi i \theta} \right\}\nonumber\\
& &+\varepsilon^2 \left\{ \hat{X}_2(0)+\hat{X}_2(2)e^{4 \pi i \theta}+\hat{X}_2(-2)e^{-4 \pi i \theta} \right\}\nonumber\\ 
& &+\varepsilon^3 \left\{ \hat{X}_3(1)e^{2 \pi i \theta}+\hat{X}_3(-1)e^{-2 \pi i \theta}+\hat{X}_3(3)e^{6 \pi i \theta}+\hat{X}_3(-3)e^{-6 \pi i \theta} \right\} + \cdots
\end{eqnarray}

When $\varepsilon$ is gradually increased from zero, the terms of higher
frequency modes get larger in order. This explains the amplification of
torus oscillation for larger $\varepsilon$, as well as the slower decay of
the Fourier coefficient with the wavenumber.


We have numerically calculated the power spectrum $P(k)=|\hat{X}(k)|^2$ for
the three types of attractors, by using the largest $2^N$ points possible out
of $F_k$ points of the attractor of functional map (\ref{MAP}).  The power
spectra $P(k)$ are plotted in Fig.\ref{F_COE}, for
$a=3.0$,$\varepsilon=0.15,0.156,0.16$, where $F_k=317811$,
$2^N=2^{18}=262144$.

\begin{figure}
\caption{Power spectrum $P(k)=|\hat{X}(k)|^2$ for $a=3.0$ and
(a)$\varepsilon=0.15$, (b)$\varepsilon=0.156$, (c)$\varepsilon=0.16$
obtained by averaging 100 different phase shifts $\theta_0$ (shift of
lattice points within the range of $1/F_k$). ($F_k=317811,
2^N=2^{18}=262144$)}
\label{F_COE}
\end{figure}

When the invariant curve is fractal, the first derivative of the Fourier
series(\ref{FS}) is expected to lose its convergence. Indeed, the spectra
$P(k)$ decrease with $k$ slowly for fractal torus. The maximum of the
envelope of $P(k)$ decays slower than $k^{-2}$, which means that
$\hat{X}(k) \sim k^{-\alpha}$, where $0< \alpha <1$.  The non-smoothness is
seen in the power spectra.  For a smooth torus the spectra decay faster
than or equal to $k^{-2}$.

It is interesting to note the analogy with the cascade process in the
turbulence.  The equation (\ref{*}) has a similar form with the Fourier
expansion of the Navier-Stokes equation, with regards to the formation of
cascade process by the term $\hat{X}(k)\hat{X}(k-k')$.  In the case of
fluid turbulence, this term brings about the Kolmogorov's energy cascade.
In contrast with the time-dependence in the
turbulence, the Fourier modes of our functional equation fall into
time-independent values at the fractal torus regime. In the fluid turbulence, 
the intermittency leads
to sporadicity in the vortex cascade, which brings about the deviation from
Kolmogorov's $5/3$ law.  In our problem this corresponds to the sporadicity
as to the extremal points.  The degree of the sporadicity changes with the
parameter $\varepsilon$, as in the intermittency effect in the turbulence,
while the sporadicity is lost at the the onset of chaos in the cascade of
the functional equation.  It will be interesting to search for a quantity
in our problem corresponding to the energy cascade in turbulence.

\section{Fractal Torus versus Smooth Torus}

\mbox{}

Let us reconsider the difference between smooth and fractal tori from the
dynamical systems viewpoint.  Of course, the largest difference between the
two types of tori is the length. The length of fractal torus is infinite,
in contrast with the smooth torus. This also leads to a difference in the
orbital instability for the dynamics of the two types of tori.  Smooth and
fractal tori are both stable against the perturbation in the
$x$-direction.  Although no exponential divergence of orbits exists for
both, they are different with each other as to the phase sensitivity.
Pikovsky and Feudel \cite{PF} have shown that two points on the SNA with
close $\theta$-values separate from each other, by introducing the
following phase sensitivity exponent: For this, note that the absolute value of the
first derivative of the orbit $|\frac{\partial x_n}{\partial \theta_n}|$
fluctuates with time and sometimes has a large burst. An arbitrarily large
burst can appear when the map is iterated for infinite time steps. By
differentiating eq. (\ref{MAP}) with respect to $\theta$, one obtains
\begin{equation}
\frac{\partial x_{n+1}}{\partial \theta} = f'(x_n) \frac{\partial x_n}{\partial \theta} 
+ \varepsilon g'(\theta)
\label{SE_EQ1}
\end{equation}
Using this equation one can compute the evolution of $\frac{\partial
 x_n}{\partial \theta}$ starting from some initial point $(x_0,\theta_0)$
 and $\frac{\partial x_0}{\partial \theta} = 0$.  The phase sensitivity by
 Pikovsky and Feudel, then, is defined as
\[
\Gamma_N = \min_{x_0,\theta_0} \max_{0 \le n \le N} 
\bigg|\frac{\partial x_n}{\partial \theta} \bigg|.
\]
This quantity must grow infinitely for SNA with some power (and
exponentially for a chaotic attractor)\cite{PF}.  In our example, this
quantity increases with the power 2.5$\sim$3 for a fractal region, as is
shown in Fig.\ref{SE}(a).  At the ``pre-fractal-torus" regime the quantity
increases up to $10^6$ steps (which corresponds to the mesh scale of
$10^{-6}$), and then saturates, as is consistent with the results of the
length in \S 2 obtained from the functional equation.

\begin{figure}
\centerline{
}
\caption{(a)$\Gamma_N$ and (b) $\Gamma_N^{(\varepsilon)}$ obtained for
$a=3.0,\omega=(\protect\sqrt{5}-1)/2$. We have chosen 100 initial points $(x_0,
\theta_0)$ randomly and iterated eqs. (18) and (19) for each initial point, 
starting from $\frac{\partial x_0}{\partial \theta} = 0$. 
$\Gamma_N$ and $\Gamma_N^{(\varepsilon)}$ are obtained
as the minimum of these 100 trajectories.  }
\label{SE}
\end{figure}

This $\Gamma_N$ corresponds to the largest derivative of the attractor
$X(\theta)$ of functional map(\ref{FUNC_MAP}) for the mesh size about $N
\times$(number of samples for $\displaystyle \min_{x_0,\theta_0}$ ). The larger the
iteration number $N$ is, the larger the length of the attractor at the mesh
is.

The second difference between smooth and fractal tori lies in the parameter
dependence of our invariant curve.  
For a fractal torus, there is
sensitivity of the curve to parameter $\varepsilon$ at least for some value
of $\theta$.  To confirm the sensitivity, we introduce the parameter
sensitivity, in the same way as $\Gamma_N$\cite{PF}: By differentiating the
first equation of (\ref{MAP}) with respect to $\varepsilon$, we obtain
\begin{equation}
\frac{\partial x_{n+1}}{\partial \varepsilon} = f'(x_n) \frac{\partial x_n}{\partial \varepsilon} 
+ g(\theta).
\label{SE_EQ2}
\end{equation}
Then we define
\[
\Gamma_N^{(\varepsilon)} = \min_{x_0,\theta_0} \max_{0 \le n \le N} 
\bigg|\frac{\partial x_n}{\partial \varepsilon} \bigg|.
\]
In Fig.\ref{SE}(b) we have plotted
this parameter sensitivity.  
Again this quantity grows with some power for the fractal torus. In our model this power
is identical with that for $\Gamma_N$, i.e., the power 2.5$\sim$3
(see Fig. \ref{SE}(b)).  This is expected since
eq.(\ref{SE_EQ1}) for $\Gamma_N^{(\varepsilon)}$ has the same form as
$\Gamma_N$ (eq.(\ref{SE_EQ2})) except the last term that is smooth.  The
power law increase of $\Gamma_N^{(\varepsilon)}$ implies that the fractal
torus always has this kind of parameter sensitivity.

Such parameter dependence is also reflected in the Lyapunov exponent of
each attractor. The sensitivity of the Lyapunov exponent to the parameter
$\varepsilon$ is relatively sharp in the region of fractal torus, while the
parameter dependence is smooth in the region of normal torus, as we have
mentioned earlier.

The third difference lies in the loss of convergence in the derivative of
$X(\theta )$ for the fractal torus.  Indeed, the Fourier expansion of
$X(\theta )$ decays faster than or equal to $1/k$ for a smooth torus, while
for the fractal one, it decays slower than $1/k$, as is discussed in the
previous section in terms of $P(k)$.

This loss of convergence corresponds to the power spectrum of the time
series of $x_n$, since the relation $\theta=\omega n+const. (\bmod 1)$
leads to a direct correspondence between $\theta$ and time.  It is shown in
\cite{Romeiras} that the power spectrum of time series is singular for SNA.
There the number of singular points whose power is larger
than a threshold decreases with some power with the threshold, while it
decays exponentially for a smooth torus.  This characteristic is another
manifestation of the loss of convergence in the Fourier modes of the
derivative.

The question remains what causes the transition from smooth to fractal
torus. One possible route to SNA is crisis between stable and unstable
invariant curves\cite{HH}. In our model, however, the unstable torus 
(continued from the unstable fixed point of the logistic map) lies far
away from the stable one and has nothing to do with the
fractalization. Hence there must be another possible mechanism for the emergence of
SNA. In the functional equation, this can be seen in the loss of
convergence in the Fourier series.  How is the loss of smoothness expressed
in terms of dynamical systems for the original 2-dimensional map?

To study this problem, we have introduced the following $F_k$-map.  If we
approximate $\omega$ by $F_{k-1}/F_k$ and start iterations from any point
$(x_0,\theta_0)$, the orbit will be periodic in $\theta$ with period $F_k$. 
($\theta_{F_k} = \theta_0 + F_k \omega = \theta_0 + F_{k-1} = \theta_0
(\bmod 1)$) Thus a composition one-dimensional map is constructed as the
$F_k$-times iterations of the original 2-dimensional map. Let us call this
composition map $F_k$-map. $F_k$-map is a function of $x$ if we fix the
initial $\theta_0$. The shape of graph of $F_k$-map depends on initial
$\theta_0$, $\varepsilon$ and $k$. Fig.\ref{Fk} gives a sequence of
examples of the $F_k$-map. The map converges to the functional map as $k$ goes to
infinity. We can infer the behavior of the functional map from the 
$F_k$-map.

In Fig.\ref{BD} we have plotted the bifurcation diagram of the $F_k$-map,
by taking $F_k=987$. ( This rather small value of $F_k$ is chosen so that
the characteristic feature of one-dimensional map is visible). Since the onset of
fractal torus and chaos can slightly differ between $\omega=F_{k-1}/F_k$ and
$\omega=(\sqrt 5-1)/2$, we have to take into account of possible shift of
the bifurcation parameter, to compare the result of $F_k$-map with that
of the functional map.

Let us look at the SNA region (0.1542 $\le \varepsilon \le$ 0.1573) in
Fig.\ref{BD}.  First, we can see sensitive dependence on $\varepsilon$, as
we have discussed earlier.  Second, we can see some points where the map
does not approach a fixed point, and instead, shows chaotic behavior. By
the definition of SNA, the $F_k$-map must have a fixed point when $k$ goes
to infinity. When $\omega$ is approximated by a rational ( say by
$F_{k-1}/F_k$ with $F_k=987$), this does not have to be true. It seems that
there always exists at least one chaotic point for some initial $\theta_0$ at
every value of $\varepsilon$ in the SNA region.  From this result we
propose the following picture of SNA viewed from the $F_k$-map.  In the SNA
region, the $F_k$-map has dense $\varepsilon$ values leading to chaotic behavior
for some $\theta_0$, whose measure goes to zero with the increase of the
mesh size $F_k$.
This kind of ``partially chaotic'' behavior may be related with the
existence of positive local Lyapunov exponent discussed by Pikovsky and
Feudel\cite{PF}.

We also note that $F_k$-map has strong sensitivity to $\theta_0$ for SNA.
Change of $F_k(x;\theta_0)$ with $\theta_0$ gets faster as $k$ goes to
infinity. This feature of $F_k$-map is consistent with the fractal nature
of torus.

\begin{figure}
\caption{Graph of $F_k$-map for $F_k=55, \theta_0=0$}
\label{Fk}
\end{figure}

\begin{figure}
\caption{Bifurcation diagram obtained from $F_k$-map for $F_k=987$. 
A sequence of $x_n$ generated by the $F_k$-map is
plotted over 100 steps after 2900 points are discarded.
The time series is plotted
for each $\varepsilon$-value. Initial condition for the
next $\varepsilon$ value is chosen from the orbit $x_n$
for the previous $\varepsilon$, while
$\varepsilon$ is incremented by 0.0001 between 0.154 and 0.1585.}
\label{BD}
\end{figure}

\section{From Fractal Torus to Chaos}
\mbox{}

What characterizes the transition from fractal torus to chaos? First, it is
clear that the functional map(\ref{FUNC_MAP}) loses its convergence.  This
is interpreted as the loss of stability of the fixed point in the functional
space. The functional map shows chaotic dynamics ( possibly having a
high-dimensional attractor) there, although the snapshot pattern of
$X_n(\theta)$ remains to be fractal with the power $0.88 \sim 1$.  Beyond
the onset of chaos, the chaotic fluctuation of
$|X_{n+1}(\theta)-X_n(\theta)|^2$ increases with $\varepsilon$.  It is
interesting to study the transition from a fixed point to chaos in the
functional space as a high-dimensional dynamical system.  As for the
bifurcation to chaos, this transition might be referred as intermittent
chaos in the functional space.

To verify this idea we have again considered the $F_k$ map, the
composite one-dimensional map, given in Fig.\ref{Fk}. From the graph a
kind of intermittent transition is seen at about $\varepsilon \sim
0.1561$. Indeed there is a tangent bifurcation from the fixed point,
by which the attractor of the $F_k$- map changes to a chaotic
one. This dependence on $\varepsilon$ does not change qualitatively
with $k$, if $k$ is large enough. Then the intermittent character of
transition from fractal torus to chaos in the functional map is
inferred by taking the limit $k \rightarrow \infty$.

As $k$ is increased, the sensitivity on $\varepsilon$ will be sharper, not
only for a fractal torus but also for a chaotic attractor.  The sensitivity
of the shape of graph leads to sensitivity of the Lyapunov exponent. This
gives an explanation to the sensitive dependence of the Lyapunov exponent on
$\varepsilon$, previously mentioned.

\section{Summary and Discussion}
\mbox{}

We have verified the existence of SNA in our quasi-periodically forced
logistic map by measuring the Lyapunov exponent, the length and the number
of extrema of attractor and power spectra. We have found that SNA exists in
a finite interval in the parameter space, and concluded that the SNA is no
other than the fractalization of torus in \cite{KK}. It should be noted
that Anishchenko et al. \cite{Ani} have also recently discussed the
fractalization of torus to chaos as SNA, by using a forced circle map.

To understand the mechanism of fractalization and transition to chaos,
we have introduced a {\it functional map (functional equation)}. The
attractor of the 2-dimensional map is represented as a fixed point of the
functional map in the functional space. In this view point, fractalization
of torus is expressed as the change of the fixed point solution of the functional map,
from smooth to non-smooth one. The transition to chaos might be regarded as
an intermittent chaos in the functional space.

We have also introduced $F_k$-map, a composition of the map over $F_k$
steps.  This $F_k$-map is a kind of cross section of the functional map.
With this map, the origin of the fractalization of torus is related with
the sensitive dependence on the initial phase $\theta_0$, while the transition to chaos is
associated with the intermittency from the fixed point in this $F_k$-map.

There has been general interest in search for a novel type of dynamics in a
forced system \cite{Roessler}, as has been pioneered by
Moser\cite{Moser}.  There the instability arising from the off-diagonal
element \cite{Courbage} in the Jacobi matrix may lead to a novel dynamical
behavior.  Such instability is also seen in the convective instability in
open flow problem\cite{Deiss}, where the instability due to the
off-diagonal element leads to a rich variety of dynamics\cite{of}.  The
strange non-chaotic attractor in our model is also due to the instability by the
off-diagonal element, as is characterized by the phase sensitivity of the
amplitude.  It is interesting to explore dynamical systems with such type
of instability in general, where the present functional map method 
may be useful.

Study of the functional equation itself is an interesting topic.  In
particular, the attractor of the functional map is high-dimensional
chaos, when the fractal torus is replaced by chaos.  The one-dimensional
string $X(\theta)$ is spatially fractal and temporally chaotic, as in the problem of
developed turbulence.  In this sense, the model may provide a novel class
of spatiotemporal chaos.  Indeed, the functional equation can be viewed as
a coupled map with a long-range coupling, by which each lattice point is
coupled with the point at a distance of $F_{k-1}$ sites.

It may be interesting to note an analogy between our functional equation and
the so called Weierstrass function.  Weierstrass function is defined as
\begin{equation}
W(\theta ) = \sum_{n=0}^{\infty} a^n \cos b^n \pi \theta \qquad 
(ab \ge 1,0<a<1)
\end{equation}
Topological dimension of Weierstrass function is one but its fractal
dimension is believed to be more than one.
The Fourier coefficient of $W(\theta)$ decreases as 
$k^{-1+\frac{\log(ab)}{\log b}}$ with the wavenumber $k$.  As Yamaguchi and Hata\cite{HY} have
demonstrated that the fractal nature of Weierstrass (and Takagi) functions
are derived from the functional equation.  Unfortunately their analysis is
not applicable to ours, since their functional equation is linear in
contrast with ours. Still some analytical studies on the functional
equation as well as the renormalization group analysis\cite{KPF} for the
perturbation series (\ref{PER_EQ}) should be of importance in future.

\vspace{.2in}

One of the authors (KK) is grateful to A.S. Pikovsky for illuminating
discussions during his stay at Potsdam.  The work is partially supported by
Grant-in-Aids for Scientific Research from the Ministry of Education,
Science, and Culture of Japan.

\end{document}